\documentclass[prl,aps,twocolumn,showpacs,floatfix]{revtex4}

\usepackage{graphicx}
\usepackage{dcolumn}
\usepackage{bm}
\newcommand{\be}{\begin{equation}}
\newcommand{\ee}{\end{equation}}
\newcommand{\bea}{\begin{eqnarray}}
\newcommand{\eea}{\end{eqnarray}}

\let\oldphi\phi
\let\phi\varphi
\let\varphi\oldphi

\begin{document}

\title{Surface elastic waves in granular media under gravity}
\author{L. Bonneau}
\author{B. Andreotti}
\author{E. Cl\'ement}
\affiliation{Laboratoire de Physique et M\'ecanique des Milieux
H\'et\'erog\`enes (UMR CNRS 7636),
10 rue Vauquelin, 75005 Paris France.}

\date{\today}

\begin{abstract}
Due to the non-linearity of Hertzian contacts, the speed of  sound $c$ in granular matter is expected to increase with pressure as $P^{1/6}$. A static layer of grains under gravity is thus stratified so that bulk waves are refracted toward the surface. We investigate wave propagation in the framework of an elastic description taking into account the main features of granular matter: non-linearity between stress and strain and existence of a yielding transition. We show in this context the existence of waves localised at the free surface --~like Rayleigh waves~-- but with a wave guide effect related to the non-linear Hertz contact.  The dispersion relationship shows a discrete number of modes which correspond to modes localized in the sagittal plane but also to transverse modes. The propagation speed of these waves is finally compared to recent measurements performed in the field, at the surface of a sand dune.  
\end{abstract}

\maketitle
\section{Introduction}
Desert animals like scorpions, use sand-born surface sound waves up to a distance of half a meter in order to localize their prey\cite{Brownell77}. On each of their legs they possess slit sensilla receptors and associated neural connections, such as to detect small phase lags and orient their killing jump very efficiently. Actually to our knowledge, the first experiments on wave propagations over sandy free surfaces were conducted in this context. Biologists like Ph.Brownell and his collaborators\cite{Brownell77} have identified surface propagation of sound and  measured a rather low speed (around $50m/s$ in the 100-500 Hz range) compatible with the animal biological capacities for signal processing\cite{Hemmen}. From the physics point of view, surface waves in an elastic material (called Raleigh waves) are traveling at speeds slightly smaller than the bulk shear waves\cite{Landau86}. Nevertheless, in the context of granular matter, the very existence of such waves and their ability to propagate over large distances is quite problematic.

Granular matter, as a collection of stiff grains under moderate confining pressure, bears inter-granular contact surfaces of a scale  much smaller than a typical grain size. In fact, the contact surfaces and the contact orientations are likely to be modified either reversibly of irreversibly under the variation of inter-granular forces. This feature is at the origin of many inherent difficulties when one wishes to determine the macroscopic constitutive properties. Moreover, the packing topology which determines the contact geometry is usually of a strongly disorderered nature \cite{Velicky} and average quantitities like deformation fields are quite subtle to define\cite{Goldenberg02}. Thus, already at the most simple level of description, involving local non-linear  elastic relations (the classical Hertz force problem\cite{H82})), a mean-field approach which identifies the local granular displacements with macroscopic deformations, is failing quantitatively\cite {Duffy57,Digby81,Walton87,Norris97,Domenico77,Makse99,Makse05}. In most cases, tests were made using sound wave bulk propagation\cite {Duffy57,Domenico77,Makse99,Jia99} under rather large confining pressures $P$ and results show a propagation velocity $c\propto P^{1/4}$ instead of the standard mean-field prediction $c\propto P^{1/6}$. Actually, Makse et al.\cite {Makse99} have shown clearly using numerical simulations that if one takes into account the effective increase of the number of contacts with pressure, the agreement is bettered (especially for compression waves). Nevertheless, an essential discrepancy still lies in the assessment of the shearing stiffness merging from local tangential contact forces\cite {Makse99,Makse05}. Note also, that from an experimental point of view, the exact origin of the discrepancy is not totally clair (\cite{Coste}). Other features such as a angular shape contacts \cite{Goddard90} or the existence of a soft layer surrounding the grains\cite{PGDG96}, can modify the propagation velocities in directions observed experimentally.
\begin{figure}[t!]
\includegraphics{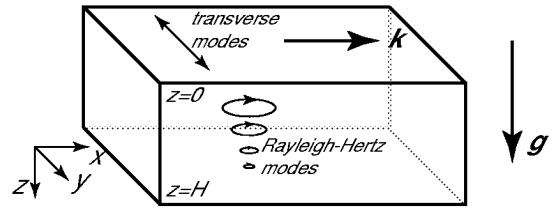}
\caption{Theoretical set-up. We consider elastic waves propagating along the $x$ direction in a cell full-filled of granular matter. Gravity is along the $z$ axis. The $y$ axis is the direction transverse to the propagation.}
\label{setup}
\end{figure}

In the limit of vanishing confining pressure very few results on sound wave propagation exist either experimentally\cite{Brownell77,Liu92} or theoretically\cite{Gussev06a}. There are a priori several difficulties due to the unilaterality of granular contacts which implies that the elastic constants are bound to vanish at zero pressure. Therefore, the elastic problem will go to a singular limit. A second order-like phase transition was recently identified in this limit, for frictionless grains around the J-point of jamming\cite{Ohern01} also displaying an anomalous density of low frequency vibrational modes \cite{Wyart}. The study was extended to packing with frictional contacts \cite{Vanecke}. In practice, if one excludes microgravity experiments under controlled confinement (yet to be performed!), this peculiar situation would naturally exists close to a free surface. A simple (mean-field) arguments, shows that under these conditions, bulk waves cannot propagate parallel to the surface. This is due to the increase of material stiffness with depth which should redirect the wave front toward the free surface (the mirage effect). Experiments conducted by Liu et al.\cite{Liu92} have shown an extreme sensitivity of the acoustic signal to minute local reorganisations, possibly due the speckle effect \cite{Jia99}, a dominant feature for probe sizes on the order of one grain.

In this paper, we address the issue of surface wave propagation in a theoretical framework of continuous non-linear elasticity. The model we use was recently introduced by Jiang and Liu \cite{JangLiu03} to describe granular constitutive  properties. In the first part of this paper, we give a short state of the art concerning the elastic description of granular media in the static phase and present the phenomenology introduced by Jiang and Liu. In a second part, we derive the equations of propagation for an isotropic compression and and uni-axial load. In the third part, we analyse the propagation of Rayleigh-like surface waves under gravity and compare the results to experimental measurements made in the field, on a dune slope.

\section{The Jiang-Liu model}
We introduce the displacement field $\rho \mathbf{U}$ and the stress tensor $\sigma_{ij}$. With our conventions, the dynamical equation reads:
\begin{equation}
\rho \mathbf{\ddot{U}}=-\mathbf{\nabla}\bar{\sigma}+\rho\mathbf{g} \nonumber
\end{equation}
The stress tensor $\sigma_{ij}$ is related to the strain tensor $u_{ij}=\frac{1}{2}\left(\frac{\partial U_{i}}{\partial x_{j}}+\frac{\partial U_{j}}{\partial x_{i}}\right)$. \\
Recently, Jiang et Liu have proposed a phenomenological formulation for the elasticity of granular states based on a simple energetic formulation : 
\begin{equation}
F_{el}=\frac{1}{2}E\delta ^{1/2}(Bu_{ll}\delta ^{2}+2u_s^{2})
\end{equation}
where $E$ is the material young's modulus, $A$ and $B$ are two dimensionless numbers. The volumic compression is :
\begin{equation}
\delta=-Tr(u_{ij})
\end{equation}
$u_{ij}^{0}=u_{ij}+\frac{\delta}{3}\delta_{ij} $ is the traceless strain tensor and
$$u_s^{2}=u_{ij}^{0}u_{ij}^{0}$$
its modulus squared. Interestingly, this elegant and compact formulation of elastic energy is able to reproduce many qualitative features observed experimentally, such as the existence of a Coulomb-like failure or stress induced anisotropy. Efforts have been made by the authors to compare quantitatively their model to the output of several experimental measurements, such as systematic triaxial tests, the response to a local load and static equilibrium in a column. The agreement was noticeable, especially in view of the minimal amount of free parameters in the model. The possibility to follow the elastic behavior up to the limit of failure is also a promising feature of the model in the context of slope stability monitored by sound waves. Note that this feature is quite original if one compares to the standard Boussinesq non-linear elasticity framework \cite{Boussinesq74}(see ref.\cite{Gussev06a}). Of course, issues like irreversible deformation fields (plasticity) are still questionable in the framework of this model; neverthelesse it provides a well defined starting point for a complete analyses and modelling of elastic vibrations. Moreover, this approach can be generalized easily with a power law not representing necessarily the Hertzian interaction. Thus, one could in principle, take into account the existence of other types of non-linear contact force-laws \cite{Goddard90,PGDG96}. In the present paper, we will limit ourselves to Hertzian interactions and we will derive sound propagation around the simplest possible reference states. 

The derivation of the stress tensor yields :
\begin{equation}
\sigma_{ij}=E \sqrt{\delta} \left(\mathcal{B}\delta\delta_{ij}-2\mathcal{A}u_{ij}^{0}+\frac{\mathcal{A}u_s^{2}\delta_{ij}}{2 \delta}\right)
\label{Liu}
\end{equation}

\subsection{Mean field determination of the compression modulus}
	Since Mindlin in the 50's, several works, focusing on sound wave bulk propagation in granular materials, have derived from local Herzian interactions, the effective elastic moduli \cite {Duffy57,Digby81,Walton87,Norris97}; for a recent and clear review, see Makse et al.\cite{Makse05} and references inside. We just recall here the mean-field results for a packing of spheres of compacity $\Phi $ and an average coordination number $Z$. First let us write the interparticular forces between two spheres of radii $R_{1}$ and $R_{2}$ and half overlap $\xi =1/2(R_{1}+R_{2}-\left| \overrightarrow{x}_{_{1}}-\overrightarrow{x}_{_{2}}\right| )$, $R=2R_{1}R_{2}/(R_{1}+R_{2})$ (if $R_{1}=R_{2}$, $R$ is the sphere radius). The normal force is :
$$F_{n}=\frac{8}{3} \frac{\mu _{g}}{1-\nu _{g}}R\left( \frac{\xi }{R}\right) ^{3/2}$$
where $\mu _{g}$ is the shear modulus and $\nu _{g}$\ the Poisson ratio.
If half the tangential separation between the sphere center is $\Delta s$, the tangential force is then :
$$\Delta F_{t}=\frac{8\mu _{g}}{2-\nu _{g}}R\left( \frac{\xi }{R}\right) ^{1/2}\Delta s$$.\\
Finally, one may obtain from mean-field granular displacement, the elastic constant values, i.e. the bulk modulus:\\ 
$$K_{MF}=\frac{1}{3\pi }\frac{\mu _{g}}{(1-\nu _{g})}\left( \Phi z\right) ^{2/3}\left( \frac{3\pi \left( 1-\nu _{g}\right) }{2\mu _{g}}P\right) ^{1/3}$$
and the shear modulus:
$$\mu _{MF}=(\frac{1}{1-\nu _{g}}+\varepsilon \frac{3}{2-\nu _{g}})\frac{\mu _{g}}{5\pi }\left( \Phi z\right) ^{2/3}\left( \frac{3\pi \left( 1-\nu _{g}\right) }{2\mu _{g}}P\right) ^{1/3}$$
with two limits (i) no sliding friction between the grains (take $\varepsilon =0$) or (ii) infinite friction between the grains (take $\varepsilon =1$).
For the macroscopic Jiang-Liu model in the isotropic case, the strain and stress tensors read :
\begin{equation}
u_{ij}=-\frac{\delta_0}{3} \delta_{ij} \,\,\,\,\,\,\,\, \sigma_{ij}=P \delta_{ij} \,\,\,\,\,\,\,\, P=E \mathcal{B} \delta_0^{3/2}
\end{equation}
For an isotropic compression, we have: $P=EB\delta _{0}^{3/2}=K_{MF}\delta _{0}$, therefore:
 $$K_{MF}\equiv EB\left(\frac{P}{EB}\right) ^{1/3}$$
 
Consequently : 
%
\begin{equation}
\mathcal{B}=\frac{\Phi z}{2^{3/2} 3\pi (1-\nu_g^2)}
\end{equation}

\subsection{Determination of the ratio $\mathcal{B}/\mathcal{A}$}
\subsubsection{Mean field}
Now we consider the Jiang-Liu model with x-z shear under isotropic compression $P$ : 
$$\sigma _{xz}=-2AE\delta _{0}^{1/2}=-2\mu _{_{MF}}u_{xz}$$ and $$P=EB\delta _{0}^{3/2}$$
Thus, 
$$\mu _{MF} \equiv EA\left( \frac{P}{EB}\right) ^{1/3}$$
Therefore :
\begin{equation}
\mathcal{A}=\frac{\Phi z}{2^{3/2} 5\pi (1+\nu_g)}\left(\frac{1}{1-\nu _{g}}+ \frac{3\varepsilon}{2-\nu _{g}}\right)
\end{equation}

If one takes $\Phi \simeq 0.6$ and $\ Z\simeq 6$.\ The bulk shear modulus for silica oxide is $\mu _{g}\simeq 30GPa$\ and the Poisson ratio is $\nu _{g}\simeq 0.2$. This gives : $EB\simeq10 GPa$ and $EA\simeq6.5 GPa$ (for $\epsilon=0$ and $EA \simeq 9 GPa$ for $\epsilon=1$. These values will be of importance when the surface sound wave velocities, found experimentally for sand, will be discussed at the end of the paper.

From the expressions above, we get:
\begin{equation}
\frac{\mathcal{B}}{\mathcal{A}}=\frac{5}{3 \left(1+ \frac{3 \varepsilon (1-\nu_g)}{2-\nu _{g}} \right)}
\end{equation}
For $\epsilon=0$, we get $\mathcal{B}/\mathcal{A}=5/3$.
For $\epsilon=1$ and $\nu _{g}=0.2$, we get $\mathcal{B}/\mathcal{A}=5/7$.

\subsubsection{Energetic argument}
	According to Jiang and Liu, we have a material instability corresponding to a Coulomb yield, for a criterion based on the free energy landscape convexity (a thermodynamic stability criterion)\cite{JangLiu03}.
\begin{equation}
\tan \theta=\sqrt{\frac{2 \mathcal{A}}{\mathcal{B}}}
\end{equation}
Thus $\mathcal{B}/\mathcal{A}=2/\tan^2 \theta_c \simeq 6$
We then notice a large descrepancy for the $B/A$ ratio between the mean-Field solution and the empirical result obtained from direct measurement of the sand-pile slope.

\section{Sound propagation under isotropic compression}
\subsection{Longitudinal waves}
Now, let us consider the propagation of longitudinal waves along the $x$ axis. As the system is homogeneous, the modes are simply Fourier modes of the form: $U e^{i(kx-\omega t)}$. We denote by $\widetilde{u_{ij}}$ the disturbance to the strain field and $\widetilde{u_{ij}}^{0}$ its traceless counterpart:
\begin{eqnarray}
\nonumber
\widetilde{u_{ij}}=i k U  \left(\begin{array}{ccc}
1& 0 & 0\\
0 & 0& 0\\
0 & 0 & 0 \end{array}\right)\,\,\,\,\,
\widetilde{u_{ij}}^{0}=i k U  \left(\begin{array}{ccc}
\frac{2}{3} & 0 & 0\\
0 & - \frac{1}{3}& 0\\
0& 0 & - \frac{1}{3} \end{array}\right)
\end{eqnarray}
The disturbance of the volumic compression is then: $\tilde{\delta}=-i k U$ and that of the modulus $u_s^{2}$ is null. The stress associated to the sound wave is:
$$\widetilde{\sigma_{ij}}=E \sqrt{\delta_0}  \left[\frac{3}{2} \mathcal{B}\tilde{\delta}\delta_{ij}-2 \mathcal{A} \widetilde{u_{ij}}^{0}\right]$$
so that the equation of motion finally reads:
\begin{equation}
- \rho \omega^2 U = - k^2 E \sqrt{\delta_0}  \left[\frac{3}{2} \mathcal{B} + \frac{4}{3} \mathcal{A}\right] U
\end{equation}
The speed of longitudinal acoustic waves is finally:
\begin{equation}
c=\gamma_\parallel \left(\frac{P}{E}\right)^{1/6}\sqrt{\frac{E}{\rho}}
\end{equation}
with $\gamma_\parallel=\left(\frac{3}{2} \mathcal{B} + \frac{4}{3} \mathcal{A}\right)^{1/2} \mathcal{B}^{-1/6}$. We thus recover the scaling of the speed of sound $c$ as $P^{1/6}$, but with an extra dependence on the coefficients $\mathcal{A}$ and $\mathcal{B}$ which themselves depends on the mean number of contacts and thus on the pressure.

\subsection{Transverse waves}
Similarly, we consider the propagation of transverse waves along the $x$ axis. The modes are still Fourier modes of the form: $V e^{i(kx-\omega t)}$. 
\begin{eqnarray}
\nonumber
\widetilde{u_{ij}}=i k V  \left(\begin{array}{ccc}
0& 1/2 & 0\\
1/2 & 0& 0\\
0 & 0 & 0 \end{array}\right)
\end{eqnarray}
The disturbance of the volumic compression $\tilde{\delta}$ and of the modulus $u_s^{2}$ are both null.
The disturbance to the stress reduces to:
$$\widetilde{\sigma_{ij}}=-2E \sqrt{\delta_0} \mathcal{A} \widetilde{u_{ij}}^{0}$$
and the equation of motion finally reads:
\begin{equation}
- \rho \omega^2 V = - k^2 E \sqrt{\delta_0} \mathcal{A} V
\end{equation}
The speed of longitudinal acoustic waves is finally:
\begin{equation}
c=\gamma_\perp \left(\frac{P}{E}\right)^{1/6}\sqrt{\frac{E}{\rho}}
\end{equation}
with $\gamma_\perp=\mathcal{A}^{1/2} \mathcal{B}^{-1/6}$. 

We thus get a prediction for the relation between the propagation speeds of longitudinal and transverse waves:
$$
\frac{\gamma_\parallel}{\gamma_\perp}=\left(\frac{3 \mathcal{B}}{2 \mathcal{A}}  + \frac{4}{3} \right)^{1/2} =\left(\frac{3}{2 \mu^2}  - \frac{7}{6} \right)^{1/2} 
$$

\begin{figure}[t!]
\includegraphics{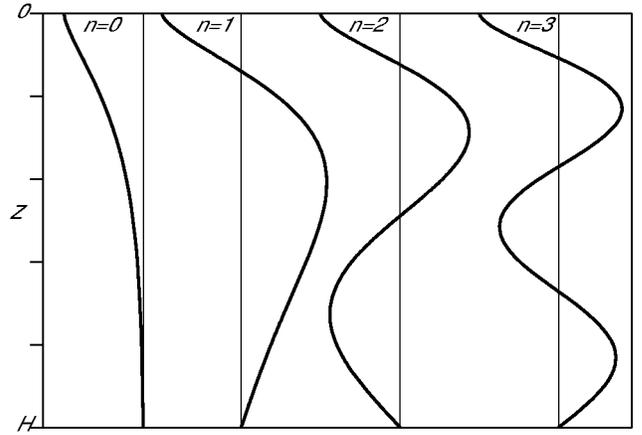}
\caption{Shape of the transverse modes for the same frequency. The modes $n>0$ are cut}
\label{ShapeTransverse}
\end{figure}
\section{Transverse surface waves under gravity}
\subsection{Equilibrium of the system}
The question is now to examine the case of a large rectangular box filled with granular matter and submitted to a vertical gravity field (fig.~\ref{setup}). As mentioned in the introduction, due to the inhomogeneous pressure field $P \propto \rho g z$, the system presents a stratification in the propagation speed $c \propto z^{1/6}$ which should also lead to a refraction toward the free surface. At first, it is clear that this mirage effect does not affect the propagation along the $z$ direction. Consequently, due to reflections also occurring at the free surface, we can expect that the box will present a series of vibration resonances. 

The first step is to compute the elastic equilibrium state of the system under gravity. We first solve the equilibrium problem, starting from a strain tensor of the form:
\begin{eqnarray}
u_{ij}=\left(\begin{array}{ccc}
0 & 0 & 0\\
0 & 0 & 0\\
0 & 0 & -\delta_0\end{array}\right)\,\,\,\,\, u_{ij}^{0}= \delta_0 \left(\begin{array}{ccc}
\frac{1}{3} & 0 & 0\\
0 & \frac{1}{3} & 0\\
0 & 0 & -\frac{2}{3}\end{array}\right)
\nonumber
\end{eqnarray}
The modulus follows as $u_s^{2}=\frac{2}{3}\delta_0^{2}$. Using the constitutive relation, we obtain the stress tensor:
\begin{eqnarray}
\sigma_{ij}=E \delta_0^{3/2} \left(\begin{array}{ccc}
\left(\mathcal{B}-\frac{\mathcal{A}}{3}\right) & 0 & 0\\
0 & \left(\mathcal{B}-\frac{\mathcal{A}}{3}\right) & 0\\
0 & 0 & \left(\mathcal{B}+\frac{5\mathcal{A}}{3}\right) \end{array}\right)
\nonumber
\end{eqnarray}
The pressure is defined from the trace of this stress tensor: $P=E \delta_0^{3/2} \left(\mathcal{B}+\frac{\mathcal{A}}{3}\right)$
The expression for the vertical stress is replaced by:
$$\sigma_{zz}=\left(\mathcal{B}+\frac{5\mathcal{A}}{3}\right) E \delta_0^{3/2}=\rho gz$$
so that the volumic compression reads:
\begin{equation}
\delta_0=-\frac{d W_0}{dz} =\left(\frac{\rho g z}{E \left(\mathcal{B}+\frac{5\mathcal{A}}{3}\right)}\right)^{2/3}
\end{equation}
where $W_0$ is the vertical displacement field. At the lower edge of the box $z=H$ (fig.~\ref{setup}), we impose that the grains remain fixed: the boundary condition is $W_0(H)=0$.
\begin{figure}[t!]
\includegraphics{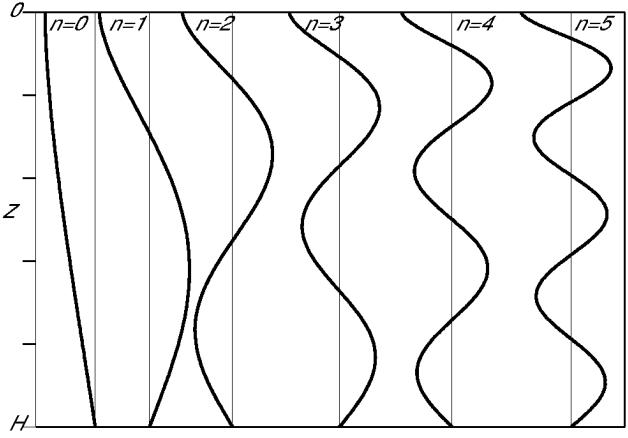}
\caption{Shape of resonant --~not propagative~-- modes.}
\label{ShapeOscillation}
\end{figure}

\subsection{Linearized equations}
Now, we investigate the existence of --~propagative~-- transverse modes localized close the surface(see ref.\cite{Gussev06b}). We consider a transverse displacement $\zeta(x,t)=\zeta_0 e^{i(kx-\omega t)}$ at the surface of the granular bed. We define the dimensionless function $V$ of the dimensionless variable $\eta=kz$:
\begin{eqnarray}
\tilde{\mathbf{U}}=\left(\begin{array}{c}
0\\
V(kz)\\
0\end{array}\right) \zeta
\end{eqnarray}
\begin{figure}[t!]
\includegraphics{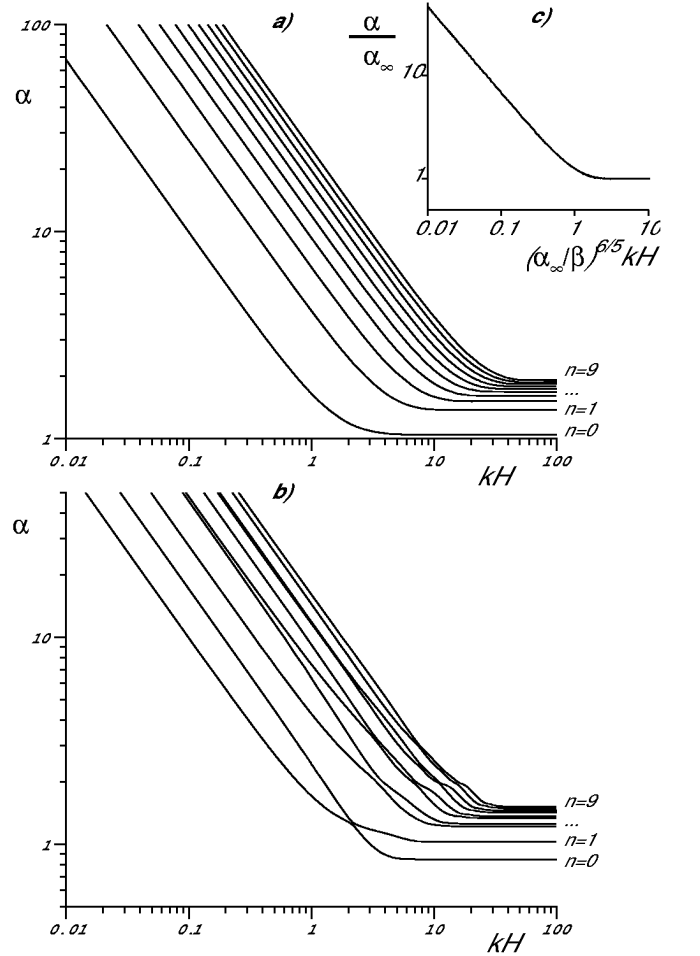}
\caption{Rescaled pulsation $\alpha$ as a function of $kH$ for Rayleigh-Hertz waves (a) and transverse surface waves (b). c) Ratio as a function of, for transverse surface modes. The different branches collapse on a single curve.}
\label{alpha_kH}
\end{figure}

Again, we linearize the strain tensor:
\begin{eqnarray}
\widetilde{u_{ij}}=k \zeta \left(\begin{array}{ccc}
0 & \frac{i V}{2} & 0\\
  \frac{i V}{2}  & 0 & \frac{V'}{2}\\
0 & \frac{V'}{2} & 0\end{array}\right)
\nonumber
\end{eqnarray}
and find no variation of the volumic compression $\tilde{\delta}=0$. The trace of $\widetilde{u_{ij}}$ is null so that $\widetilde{u_{ij}}^{0}=\widetilde{u_{ij}}$, and then $\widetilde{u_s}^{2}=2 u_{ij}\widetilde{u_{ij}}$ vanishes. Finally we obtain :
\begin{eqnarray}
\widetilde{\sigma}_{ij}=-2 \mathcal{A} E \sqrt{\delta_{0}}  \widetilde{u_{ij}}=- \mathcal{A} E \sqrt{\delta_{0}} k \zeta \left(\begin{array}{ccc}
0 & i V & 0\\
  i V  & 0 & V'\\
0 & V' & 0\end{array}\right)
\nonumber
\end{eqnarray}

Again, from the equation of motion along $y$, we get a dispersion relation within a constant $\alpha$:
\begin{equation}
\omega=  \alpha  \mathcal{A}^{1/2} (\mathcal{B}+5\mathcal{A}/3)^{-1/6} \left(\frac{E}{\rho}\right)^{1/3} g^{1/6} k^{5/6} 
\end{equation}
The shape of the mode is given by the equation:
\begin{equation}
\alpha^{2} V = \eta^{1/3} V -\left(\eta^{1/3} V' \right)'
\end{equation}

\subsection{Resolution}
We decompose the problem into two equations:
\begin{eqnarray}
\begin{array}{l}
V'= \eta^{-1/3} S\\
S'=(\eta^{1/3}-\alpha^{2})V\\
\end{array}\nonumber
\end{eqnarray}
with the boundary conditions coming from the zero stress condition at the free surface and the definition of $\zeta$: $V(0)=1$, $V'(0)=0$, $V(kH)=0$.

For infinite depth, the system presents a discrete number of modes which may be associated to the stratification. The figure \ref{alphaHetaHTrans} shows the variation of $\alpha_\infty$ with $n$ together with a fit by $(\alpha_0+\delta \alpha n)^{1/6}$. We have computed systematically the ratio $\alpha/\alpha_\infty$ as a function of $kH$. For small $kH$, we find that $\alpha$ decreases like $(kH)^{-5/6}$. For $kH$ much larger than $1$, $\alpha$ tends to $\alpha_\infty$. The curves for the different modes collapse when $kH$ is properly rescaled.

\subsection{Wave-guide cut-off: horizontal resonant modes}
The height of the box is the only relevant length scale of the problem. We introduce accordingly the coordinate  $\eta=z/H$. 
We consider modes of vibration whose disturbed displacement field is horizontal, of the form $U(z/H)e^{i\omega t}$. The disturbed strain field reads:
\begin{eqnarray}
\widetilde{u_{ij}}= \left(\begin{array}{ccc}
0 & 0 &U'(\eta)/2H\\
 0 & 0 & 0\\
U'(\eta)/2H & 0 & 0\end{array}\right)
\nonumber
\end{eqnarray}
and find no variation of the volumic compression $\tilde{\delta}=0$. The trace of $\widetilde{u_{ij}}$ is null so that $\widetilde{u_{ij}}^{0}=\widetilde{u_{ij}}$, and $\widetilde{u_s}^{2}=2 u_{ij}\widetilde{u_{ij}}$ vanishes. We end with $\widetilde{\sigma}_{ij}=-2 \mathcal{A} E \sqrt{\delta_{0}}  \widetilde{u_{ij}}$. The disturbed vertical stress deduces as:
$$\widetilde{\sigma_{zx}}=- \mathcal{A} E \sqrt{\delta_{0}} U'/H$$
and leads to the equation of motion:
\begin{equation}
\left(\eta^{1/3}U'(\eta)\right)'+\beta^2 U(\eta)=0
\label{modevibshape}
\end{equation}
where $\beta$ is the rescaled pulsation defined by:
\begin{equation}
\omega= \beta \mathcal{A}^{1/2} \left(\mathcal{B}+\frac{5\mathcal{A}}{3}\right)^{-1/6}\left(\frac{E}{\rho}\right)^{1/3} g^{1/6} H^{-5/6} 
\end{equation}
Remarkably, the resonant frequencies are found to scale as $H^{-5/6}$ and not $H^{-1}$ as for a non dispersive medium.

The solution of the ordinary equation (\ref{modevibshape}) involves the Bessel function of first kind $J_{2/5}$ and is of the form:
\begin{equation}
U(\eta)=\eta^{2/3} J_{-2/5}\left(\frac{6\beta}{5} \eta^{5/6}\right)
\end{equation}
The value of $\beta$ is selected by the zero displacement condition at the bottom edge of the box, $U(1)=0$, which simplifies into $J_{2/5}(6\beta/5)=0$. There is thus a discrete number of modes, labelled by $n$, whose rescaled frequency is:
\begin{equation}
\beta=\left(\frac{\pi}{2}\right)^{6/5}+\frac{5 \pi n}{6}
\end{equation}

\section{Rayleigh-Hertz waves under gravity}
\subsection{Linearized equations}
Now we investigate the existence of propagative modes analogous to the Rayleigh modes i.e. whose deformations are in the sagital plane $(x,z)$. We consider the deformation of the free surface to be of the form $\zeta(x,t)=\zeta_0 e^{i(kx-\omega t)}$. The displacement field is defined trough two dimensionless functions $U$ and $W$ of the dimensionless variable $\eta=kz$:
\begin{eqnarray}
\tilde{\mathbf{U}}=\left(\begin{array}{c}
i U(kz)\\
0\\
W(kz)\end{array}\right) \zeta
\nonumber
\end{eqnarray}
The disturbance to the strain tensor reads:
\begin{eqnarray}
\widetilde{u_{ij}}=k \zeta \left(\begin{array}{ccc}
-U & 0 & \frac{i (U'+W)}{2}\\
 0 & 0 & 0\\
\frac{i (U'+W)}{2}& 0 & W'\end{array}\right)
\nonumber
\end{eqnarray}
and is traceless counterpart is:
\begin{eqnarray}
\widetilde{u_{ij}}^{0}=k \zeta  \left(\begin{array}{ccc}
-\frac{2 U}{3} - \frac{W'}{3}& 0 & \frac{i (U'+W)}{2}\\
0 & \frac{U}{3} - \frac{W'}{3}& 0\\
\frac{i (U'+W)}{2}& 0 & \frac{2W'}{3}+\frac{U}{3} \end{array}\right)
\nonumber
\end{eqnarray}
The disturbance to the volumic compression is $\tilde{\delta}=k \zeta (U-W')$. That of the modulus $u_s^{2}$ reads:
$$\widetilde{u_s}^{2}=2 u_{ij}^0\widetilde{u_{ij}^0}=-\frac{2 \delta_{0}}{3}\left(U+2W'\right) k\zeta$$

The disturbance to the stress tensor may be formally expressed as:
\begin{eqnarray}
\widetilde{\sigma}_{ij}=\sqrt{\delta_0} E \left[\frac{3}{2} \mathcal{B}\tilde{\delta}\delta_{ij}-\mathcal{A} \left(\begin{array}{ccc}
\frac{1}{3} & 0 & 0\\
0 & \frac{1}{3} & 0\\
0 & 0 & -\frac{2}{3}\end{array}\right) \tilde{\delta} \right.\nonumber \\
\left.-2\mathcal{A} \widetilde{u_{ij}}^{0}-\frac{\mathcal{A}}{6} \tilde{\delta}\delta_{ij}+\frac{\mathcal{A}}{2 \delta_{0}}\widetilde{u_s}^{2}\delta_{ij}\right]
\nonumber
\end{eqnarray}
which gives, after simplification, the following expressions for its components:
\begin{eqnarray}
\begin{array}{l}
\widetilde{\sigma_{xx}}=E\sqrt{\delta_{0}}\left[ \left( \frac{\mathcal{A}}{2}+\frac{3\mathcal{B}}{2}\right)U+ \left(\frac{\mathcal{A}}{2}-\frac{3\mathcal{B}}{2}\right)W' \right] k\zeta\\
\widetilde{\sigma_{zz}}=-E\sqrt{\delta_{0}}\left[ \left( \frac{\mathcal{A}}{2}-\frac{3\mathcal{B}}{2}\right)U+\left(\frac{5\mathcal{A}}{2}+\frac{3\mathcal{B}}{2}\right)W'\right] k\zeta\\
\widetilde{\sigma_{xz}}=\widetilde{\sigma_{zx}}=-i \mathcal{A}E\sqrt{\delta_{0}}\left(U'+W\right) k\zeta\\
\end{array}
\nonumber
\end{eqnarray}
For the sake of simplicity, we rescale the stress tensor, introducing the functions $S_{xz}(kz)$, $S_{zz}(kz)$, $S_{xx}(kz)$:
\begin{eqnarray}
\widetilde{\sigma_{xz}}&=&i (Ek)^{2/3}(\rho g)^{1/3} \mathcal{A}(\mathcal{B}+5\mathcal{A}/3)^{-1/3} \zeta S_{xz}(kz) \nonumber\\
\widetilde{\sigma_{zz}}&=&(Ek)^{2/3}(\rho g)^{1/3} \mathcal{A}(\mathcal{B}+5\mathcal{A}/3)^{-1/3} \zeta S_{zz}(kz)\nonumber\\
\widetilde{\sigma_{xx}}&=&(Ek)^{2/3}(\rho g)^{1/3} \mathcal{A}(\mathcal{B}+5\mathcal{A}/3)^{-1/3} \zeta S_{xx}(kz)
\end{eqnarray}
The previous expressions then simplify into:
\begin{eqnarray}
S_{xx}(\eta)&=\eta^{1/3}  &\left[ \left( \frac{1}{2}+\frac{3 \mathcal{B}}{2 \mathcal{A}}\right)U(\eta)+ \left(\frac{1}{2}-\frac{3\mathcal{B}}{2\mathcal{A}}\right)W'(\eta) \right] \nonumber\\
S_{zz}(\eta)&=- \eta^{1/3}  &\left[ \left( \frac{1}{2}-\frac{3\mathcal{B}}{2\mathcal{A}}\right)U(\eta)+\left(\frac{5}{2}+\frac{3\mathcal{B}}{2\mathcal{A}}\right)W'(\eta)\right] \nonumber\\
S_{xz}(\eta)&=- \eta^{1/3} &\left[U'(\eta)+W(\eta)\right]
\end{eqnarray}
\begin{figure}[t!]
\includegraphics{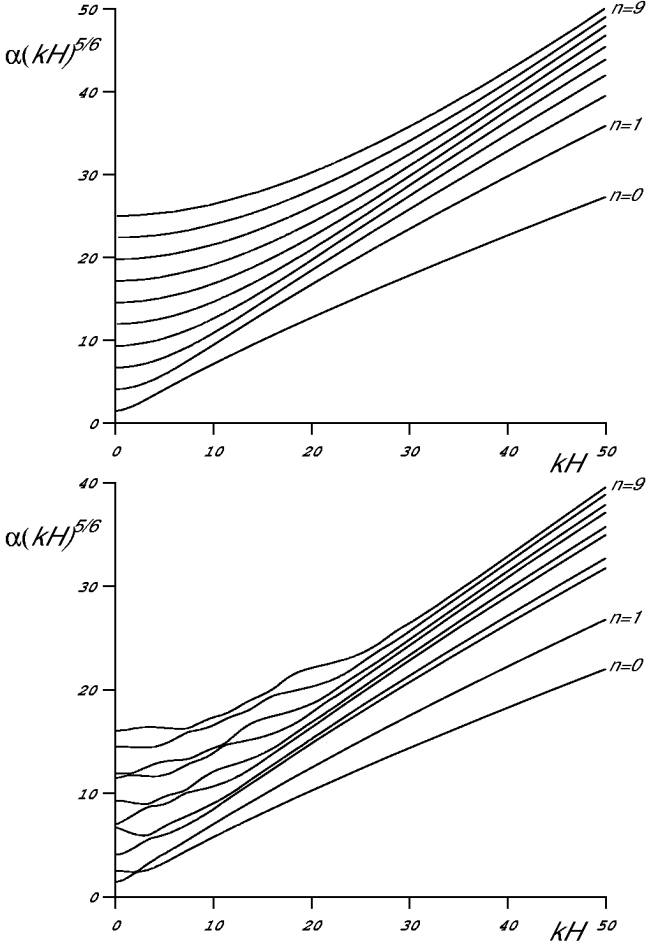}
\caption{Dispersion relation of longitudinal and transverse waves. }
\label{Dispersion_longi_trans}
\end{figure}

From the equations of motion, we get the dispersion relationship within a constant $\alpha$:
\begin{equation}
\omega= \alpha \mathcal{A}^{1/2} (\mathcal{B}+5\mathcal{A}/3)^{-1/6} \left(\frac{E}{\rho}\right)^{1/3} g^{1/6} k^{5/6} 
\end{equation}
as well as a set of equations governing the shape of the modes:
\begin{eqnarray}
\alpha^{2} U =S_{xx}+S_{xz}'\\
\alpha^{2} W =-S_{xz}+S_{zz}'
\end{eqnarray}

\subsection{Resolution}
In order to solve the boundary condition problem, we rewrite the equations as a set of $4$ --~linear~-- ordinary differential equations:
\begin{eqnarray}
U'&=&-W-\eta^{-1/3} S_{xz}\nonumber\\
W'&=&-\frac{1}{5\mathcal{A}+3\mathcal{B}} \left[\left(\mathcal{A}- 3\mathcal{B}\right)U+2\mathcal{A} \eta^{-1/3} S_{zz}\right]\nonumber\\
S_{xz}'&=&\alpha^{2} U - S_{xx}\nonumber\\
S_{zz}'&=&\alpha^{2} W+S_{xz}\nonumber\\
S_{xx}&=&\frac{1}{5\mathcal{A}+3\mathcal{B}} \left[\left(3\mathcal{B}-\mathcal{A}\right)S_{zz}+2\left(\mathcal{A}+6\mathcal{B}\right) \eta^{1/3} U\right]
\nonumber
\end{eqnarray}
with the boundary conditions: $U(kH)=0$, $W(0)=1$, $W(kH)=0$, $S_{xz}(0)=0$ and $S_{zz}(0)=0$. The solution $[U,W,S_{xz},S_{zz}](\eta)$ is obtained by superposition of the solutions obtained, starting from the initial conditions $[0,1,0,0]$ and $[1,0,0,0]$. The value of $\alpha$ is then tuned to get the cancellation of both $U(kH)$ and $W(kH)=0$.

\begin{figure}[t!]
\includegraphics{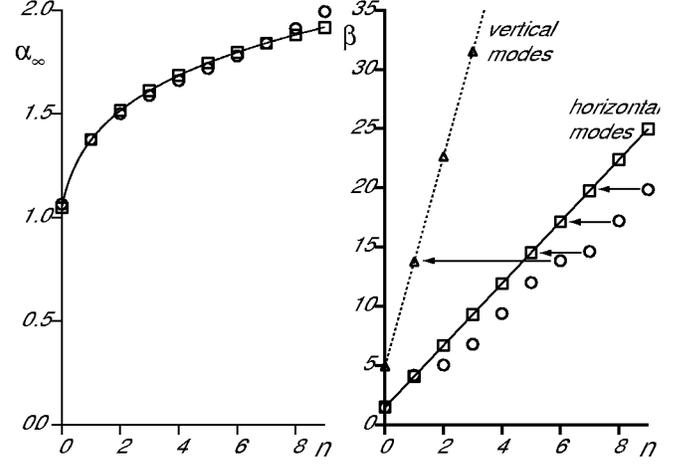}
\caption{{\bf a} Asymptotic value of the rescaled frequency $\alpha$ in the limit of large $kH$, as a function of the mode number $n$. {\bf b} Asymptotic value of the rescaled frequency $\beta$ in the limit of small $kH$. }
\label{alphaHetaHTrans}
\end{figure}

\begin{figure}[t!]
\includegraphics{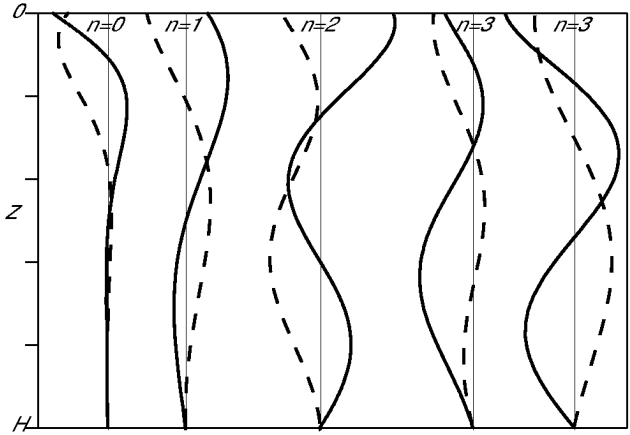}
\caption{Shape of the Rayleigh-Hertz modes for a same frequency. The modes $n>0$ are cut}
\label{ShapeRayleighHertz}
\end{figure}

The central result of this part is that contrarily to the classical Rayleigh waves, the propagation of these "Rayleigh Hertz" surface waves presents a discrete number of modes that may be associated to the stratification effect. The shape of the modes excited at a given pulsation $\omega$ is displayed on fig.~\ref{ShapeRayleighHertz}. For infinite depth i.e. in the limit of infinite $kH$, the rescaled pulsation $\alpha$ tends toward a constant $\alpha_\infty$ whose dependence on the mode number $n$ is shown on fig.~\ref{alphaHetaHTrans}. There is a general trend of $\alpha_\infty$ to increase as $n^{1/6}$ but the values are not regularly spaced.

For small $kH$, we find the usual wave-guide cut-off  as $\omega$ tends to a constant. In terms of $\alpha$, this corresponds to an asymptotic behavior of the form:
\begin{equation}
\alpha=\beta (kH)^{-5/6}
\end{equation}
So, in the limit of wavelength large compared to the cell depth, the pulsation tends to the value
\begin{equation}
\omega= \beta \mathcal{A}^{1/2} (\mathcal{B}+5\mathcal{A}/3)^{-1/6} \left(\frac{E}{\rho}\right)^{1/3} g^{1/6} H^{-5/6} 
\end{equation}
The value of $\beta$ as a function of $n$ is displayed on fig.~\ref{alphaHetaHTrans}. Although it is not a regularly increasing function of $n$, $\beta$ follows a general trend of being linearly dependent on $n$.

\subsection{Wave-guide cut-off: vertical resonant modes}
We now consider modes of vibration whose disturbed displacement field is vertical and of the form $W(z/H)e^{i\omega t}$. Then the disturbed strain field reads:
\begin{eqnarray}
\widetilde{u_{ij}}= \left(\begin{array}{ccc}
0 & 0 &0\\
 0 & 0 & 0\\
0 & 0 & W'(\eta)/H\end{array}\right) \,\,\,\,\,\, 
\widetilde{u_{ij}}^{0} = \left(\begin{array}{ccc}
- \frac{W'}{3H}& 0 & 0 \\
0 & - \frac{W'}{3H} & 0 \\
0 & 0 & \frac{2W'}{3H} \end{array}\right)
\nonumber
\end{eqnarray}
from which we deduce the disturbed volumic compression $\tilde{\delta}=-W'/H$ and the disturbed modulus
$$\widetilde{u_s}^{2}=2 u_{ij}^0\widetilde{u_{ij}^0}=-\frac{4 \delta_{0} W'}{3 H}$$
Then the disturbed vertical stress is deduced as:
$$\widetilde{\sigma_{zz}}=-\frac{E\sqrt{\delta_{0}}}{2H} \left(5\mathcal{A}+3\mathcal{B}\right)W'$$
and leads to the same equation of motion as previously
\begin{equation}
\left(\eta^{1/3}W'(\eta)\right)'+\beta^2 W(\eta)=0
\label{modevibshape}
\end{equation}
but where $\beta$ is now defined by:
\begin{equation}
\omega= \beta \sqrt{\frac{3}{2}} \left(\frac{\left(\mathcal{B}+\frac{5\mathcal{A}}{3}\right) E }{\rho}\right)^{1/3} g^{1/6} H^{-5/6} 
\end{equation}

\section{Experimental measurements and Discussion}
\begin{figure}[t]
\includegraphics{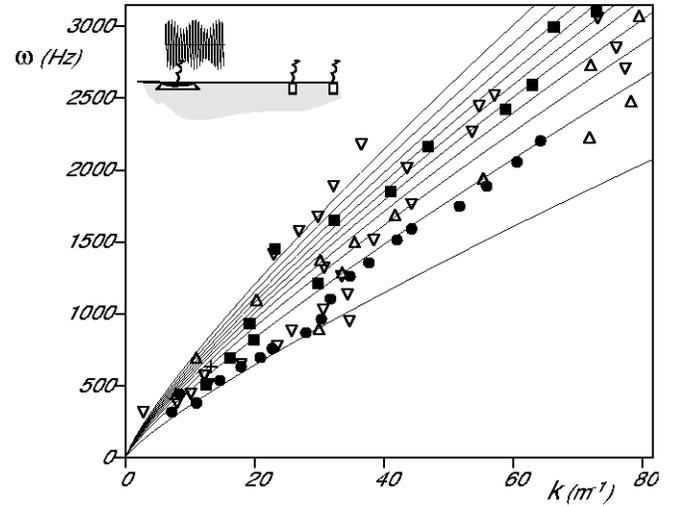}
\caption{Dispersion relation of surface elastic waves on a dune slip face. Transducers distant by $5{\rm~cm}$ (down triangle), $15{\rm~cm}$ (up triangle), $25{\rm~cm}$ (square) and $42{\rm~cm}$ (circle). The solid lines corresponds to the saggittal Rayleigh-Herz modes (see equ. ) with a material stiffness adjusted to match the lowest $n=0$ propagating mode.} 
\label{Dispersion}
\end{figure}

	In conclusion, we performed a derivation of sound-wave propagation modes in the context of the Jiang-Liu model of granular non-linear elasticity. First, the longitudinal and transverse sound velocities for and isotropic compression were calculated. Morerover, we derived the propagating and non propagating modes for a granular packing with a free surface under a vertical gravity field. We show that the gravity field producing a stratification of the material stiffnes is responsible for a channeling effect of the acoustic waves. Hence, we find two surface propagation modes in the transverse and saggital directions. The last one, we call Rayleigh-Hertz waves. 
	Now we present the results of experimental measurements performed on a sand dune using localized surface excitation and already reported in reference \cite{Andreotti04}. Sinusoidal signals, recorded on a tape, were played through an amplified loud-speaker that was kept at the surface of the avalanche slip face of a singing dune (top schematic of fig.~\ref{Dispersion}).

	As the mechanical transmission of the sound generator with the sand bed is non-linear, the signal is distorted and presents harmonics. This imperfection was used to derive several points of the dispersion relation from each recording. From the phase between the Fourier components of the signals of two accelerometers aligned with the loud-speaker, the wavelengths $\lambda$ associated to each frequency $f$ were obtained. In order to unwrap the phase, the amplitude of the signals was slightly modulated at low frequency. On fig.~\ref{Dispersion}, we present the collection of all measurements that establish the dispersion relation. 
	Actually, when replaced in the context of Rayleigh-Herz surface waves, the present framework sheds a new light on the apparent large scatter of the data. Indeed, we can superpose  on the experimental results, the dispersion relations found for the sagittal vibrations. On the figure, the solid lines correspond to the dispersion relations : $f= c_n k^{5/6}$, with $c_{n}=\alpha _{\infty }(n)A^{1/2}(B+5A/3)^{-1/6}(E/\rho )^{1/3}g^{1/6}$  corresponding to the $n^{th}$ mode prefactor derived for an infinite depth (see fig.~\ref{alphaHetaHTrans}a for $\alpha _{\infty }(n)$ values). By best fit, the $c_0$ value was adjusted to match the graviest mode and we obtained $c_0\simeq 50 USI$. Note that the agreement for $n=0,1$ and $2$ is quite good. If we compare this result to the values derived from the mean-field estimation $EB=10GPa$ and the $B/A=6$ derived from the sand-pile slope, we get  $c_0\simeq 105 USI$. Which means that the actual surface waves speeds measured in these conditions are still slow by a factor $2$ when evaluated from the simple arguments presented above (or alternatively would correspond to an overestimation of $A$ by a factor $4$). Thus, in spite of this semi-quantitative agreement, there are still many question concerning the direct evidence of the propagation modes derived in this context. This is an ongoing work in our laboratory and we leave these questions for a future report.

\end{document}